\documentclass[12pt]{article}
\usepackage{amssymb,amsmath,amscd,amsthm}

\newtheorem{theorem}{Theorem}[section]
\newtheorem{corollary}[theorem]{Corollary}
\newtheorem{assumption}[theorem]{Assumption}

\newtheorem{proposition}[theorem]{Proposition}
\newtheorem{definition}[theorem]{Definition}

\date{}

\begin{document}

\title{Ellipticity in the interior transmission problem in anisotropic media}

\author{ E.Lakshtanov\thanks{Department of Mathematics, Aveiro University, Aveiro 3810, Portugal.  This work was supported by {\it FEDER} founds through {\it COMPETE}--Operational Programme Factors of Competitiveness (``Programa Operacional Factores de Competitividade'') and by Portuguese founds through the {\it Center for Research and Development in Mathematics and Applications} (University of Aveiro) and the Portuguese Foundation for Science and Technology (``FCT--Fund\c{c}\~{a}o para a Ci\^{e}ncia e a Tecnologia''), within project PEst-C/MAT/UI4106/2011 with COMPETE number FCOMP-01-0124-FEDER-022690, and by the FCT research project
PTDC/MAT/113470/2009 (lakshtanov@rambler.ru).} \and
 B.Vainberg\thanks{Department
of Mathematics and Statistics, University of North Carolina,
Charlotte, NC 28223, USA. The work was partially supported  by the NSF grant DMS-1008132 (brvainbe@uncc.edu).}}

\maketitle

\begin{abstract}
The paper concerns the discreteness of the eigenvalues and the solvability of the interior transmission problem for anisotropic media. Conditions for the ellipticity of the problem are written explicitly, and it is shown that they do not guarantee the discreteness of the eigenvalues. Some simple sufficient conditions for the discreteness and solvability are found. They are expressed in terms of the values of the anisotropy matrix and the refraction index at the boundary of the domain. The discreteness of the eigenvalues and the solvability of the interior transmission problem are shown if a small perturbation is applied to the refraction index.

\end{abstract}



\section{Introduction}
The interior transmission eigenvalue problem in isotropic media was first introduced in 1986 by Kirsch \cite{kirsch0} in connection with the properties of the far field operator and an inverse scattering problem for the reduced wave equation. The further results on this problem were obtained by Colton
and Monk \cite{base}, see the review \cite{cps} for more references. A study of the same problem for anisotropic media was initiated in \cite{col}. One of the important  questions in the area is whether the set of the eigenvalues of the interior transmission problem is discrete. Conditions for the discreteness of the eigenvalues were extensively studied. The properties of the media are described by a matrix $A(x)=A^t(x)$ and a function $n(x)$  which determine the coefficients of the corresponding equation in a bounded domain $\mathcal O$. This paper concerns the anisotropic case ($A\neq a(x)I$ were $a$ is a function, $I$ is the identity matrix).  The discreteness in the anisotropic case was proved under certain conditions on $A,~n$ in the whole domain $\mathcal O$: $A \neq I, n \neq 1 $ at each point $x\in \overline{\mathcal O}$ or $A>I, n \equiv 1$ for all $x\in \overline{\mathcal O}$ (\cite{haddar3},\cite{haddar2},\cite{kirsch},\cite{kirsch2}).  Then it was shown (see \cite{haddar1}) that it is sufficient to have inequalities $A>I,n>1$ or $0<A<I, n<1$ only at the boundary $\partial\mathcal O$.

This paper contains much more general conditions which guarantee the discreteness of the interior transmission eigenvalues and the solvability of the interior transmission problem for anisotropic media. If $A$ and $n$ are real valued, these conditions are imposed on the values of $A,n$ only at the boundary of the domain, and the conditions are much less restrictive than the ones mentioned above. The results are based on the theory of elliptic problems and parameter-elliptic problems. The latter problems are also called the elliptic problems with a parameter, see \cite{agr},\cite{grubb} and references there; they are not simply elliptic problems depending on a parameter. For example, $\Delta-k^2,~k>0,$ is a parameter-elliptic equation, and $\Delta+k^2,~k>0,$ is not. Note that the ellipticity of the interior transmission problem alone does not lead automatically to the discreteness of the eigenvalues (counterexamples are provided in the paper), and some additional assumptions are needed.

We describe now our assumptions in the case of real valued $A$ (the general complex $A$ are considered in the paper). As usual, we assume that the equations under consideration are elliptic. This implies that all the eigenvalues of $A$ have the same sign. The ellipticity of a boundary value problem requires the ellipticity of the equations and a certain relation between the equations and the boundary conditions. The latter is called the Shapiro-Lopatinskii condition. We will show that the Shapiro-Lopatinskii condition for the two-dimensional ($d=2$) interior transmission problem is equivalent to the following relation at the boundary of the domain: det$A(x^0)\neq 1,~x^0\in \partial\mathcal O$. While this condition does not guarantee the discreteness of the eigenvalues, it will be shown that the discreteness takes place for generic potentials. Namely, it is true for potentials $cn(x)$ with arbitrary complex valued $n$ when the complex constant $c$ takes any value except possibly from a finite set $\{c_j\},1\leq j\leq N.$

Sufficient conditions for the discreteness will be given for an individual $n(x)$ (when $c=1$). In the two-dimensional case, we assume that the following relations hold at the boundary: det$A(x^0)\neq 1,~x^0\in \partial\mathcal O,$ and $an(x^0)\neq 1,~x^0\in \partial\mathcal O$. Here $a=\nu\cdot A\nu$ and $\nu$ is the unit normal to $\partial\mathcal O$. Then the interior transmission eigenvalues form a discrete set if $n(x)\neq 0,~x\in \overline{\mathcal O}$, and $n$ is real or $|{\rm Im}~ n|\geq \varepsilon>0$. In the three-dimensional case, we impose a somewhat more complicated inequality on the coefficients of the matrix $A$ at the boundary. For example, one can consider real valued elliptic $A$ such that $A > I$ or $0<A<I$
at the boundary if $n$ satisfies the same properties as in the two-dimensional case $d=2$. One also may consider matrices $A$ with some eigenvalues larger than one and some smaller than one.  The simplest example is given by a matrix which is diagonal at the boundary of the domain in a basis which consists of two vectors tangential to the boundary and of the normal $\nu$. If $\lambda_1, \lambda_2, \lambda$ are the diagonal elements of the matrix and $\lambda_1\leq\lambda_2$, then our conditions hold when $1/\lambda\notin [\lambda_1,\lambda_2]$.


\section{Main results}

Let $\mathcal O\in R^d$ be an open bounded domain with $C^2$ boundary $\partial O$ and the outward normal $\nu$. We will be mostly concerned with the cases $d=2,3$, but all the results below can be automatically carried over to any dimension $d$. Given $f\in H^{3/2}(\partial \mathcal O),  ~~g\in H^{1/2}(\partial \mathcal O)$, the \textit{interior transmission problem (ITP)} \cite{haddar2} consists of funding functions $u,v\in H^2(\mathcal O)$ satisfying the equations
\begin{equation}\label{Anone}
\begin{array}{l}
\Delta u + k^2 u =0, \quad x \in \mathcal O ,\\
\nabla A \nabla v + k^2 n(x)v =0, \quad x \in \mathcal O,
\end{array}
\end{equation}
and the boundary conditions
\begin{equation}\label{Antwo}
\begin{array}{l}
u-v=f, \quad x \in \partial \mathcal O, \\
\frac{\partial u}{\partial \nu} - \frac{\partial v}{\partial \nu_A}=g, \quad x \in \partial \mathcal O.
\end{array}
\end{equation}
Here $H^{2}(\mathcal O), ~H^{s}(\partial \mathcal O)$ are Sobolev spaces, $A(x),~x\in \overline{\mathcal O}$ is a smooth symmetric ($A=A^t$) elliptic matrix with complex valued entries, $n(x)$ is  a smooth function, and the co-normal derivative is defined as follows
$$
\frac{\partial } {\partial \nu_A}v =\nu \cdot A \nabla v.
$$
The values of $k \in \mathbb C$ for which the homogeneous problem (\ref{Anone}), (\ref{Antwo}) has a non-trivial solution are called the \textit{interior transmission eigenvalues}.

\begin{definition}\label{d1}
Recall that the ellipticity of the matrix $A$ in dimension $d\geq3$ means that the quadratic form with the matrix $A$  is not degenerate, i.e., $\xi\cdot A(x)\xi\neq 0$ for $0\neq \xi\in R^d$.
\end{definition}
The following property of the elliptic matrices is important. Given two linearly independent vectors $0\neq \xi^{(1)},\xi^{(2)}\in R^{d}$, let us solve the quadratic equation
\begin{equation}\label{ckv}
(\xi^{(1)}+\lambda\xi^{(2)})\cdot A(x)(\xi^{(1)}+\lambda\xi^{(2)})= 0
\end{equation}
with respect to $\lambda$. Then one of the roots will have a positive imaginary part and another root will have a negative imaginary part. This property is more restrictive than the non-degeneracy of the quadratic form when the dimension $d=2$, and it is usually included in the definition of the ellipticity:
\begin{definition}\label{d2}
When $d=2$, a matrix $A$ is elliptic if, for any two linearly independent vectors $0\neq \xi^{(1)},\xi^{(2)}\in R^{2}$, one of the roots of the quadratic equation (\ref{ckv}) has a positive imaginary part and another root has a  negative imaginary part.
\end{definition}
The ellipticity of the equations (\ref{Anone}) is not enough for the whole boundary value problem to be elliptic. An additional condition (Shapiro-Lopatinskii condition) is needed for the problem (\ref{Anone}), (\ref{Antwo}) to be elliptic. The latter condition depends on the coefficients of the equations at the boundary and on the boundary operators. In order to state this condition, let us fix an arbitrary point $x^0\in \partial \mathcal O $ and choose a new orthonormal basis $\{e_j\},~1\leq j\leq d,$ centered at the point $x^0$ with $e_d=\nu$, where $\nu$ is the normal to the boundary at the point $x^0$.  The vectors $e_1,...,e_{d-1}$ belong to the tangent plane to $\partial \mathcal O $ at the point $x_0$. Let $y$ be the local coordinates defined by the basis $\{e_j\}$, and let $C=C(x^0)$ be the transfer matrix, i.e., $y=C(x-x^0)$.

We fix the point $x=x^0$ in  equations (\ref{Anone}), (\ref{Antwo}) and rewrite the problem in the local coordinates $y$. Then we get the following problem with constant coefficients in the half space $y_d>0:$
\begin{equation}\label{Anone1}
\begin{array}{l}
\Delta_y u + k^2 u =0, \quad y_d>0 ,\\
 \nabla_y \widetilde{A} \nabla_y v + k^2 n(x^0)v =0,  \quad y_d>0,
\end{array}
\end{equation}
\begin{equation}\label{Antwo1}
\begin{array}{l}
u-v=f, \quad y_d=0, \\
\frac{\partial u}{\partial {y_d}} - \frac{\partial v}{\partial n_{\widetilde{A}}}=g, \quad y_d=0.
\end{array}
\end{equation}
Here
$$
\widetilde{A}=\widetilde{A}(x^0)=CA(x^0)C^*.
$$
The entries of the matrix $\widetilde{A}=(a_{i,j})$ are equal to $a_{i,j}=e_j\cdot A(x^0)e_i$. The co-normal derivative in the boundary condition equals $e_d\cdot \widetilde{A}\nabla_y.$

It will be shown below that the following assumptions coincide with the Shapiro-Lopatinskii condition for the ellipticity of the problem (\ref{Anone}), (\ref{Antwo}).
\begin{assumption}\label{a3} If $d=2$, then we assume that at least one of the following two conditions holds at each point $x^0\in \partial \mathcal O$: {\rm det}$A(x^0)\neq 1$ or ${\rm Re}~a_{2,2}<0$.
\end{assumption}
If $d=3$, then consider the following $2\times2$ matrix
\begin{equation}\label{mb}
B=\left(\begin{array}{l}
\sum_{i,j=1}^2a_{i,j}\tau_i\tau_j,~~~a_{1,3}\tau_1+a_{2,3}\tau_2 \\
a_{1,3}\tau_1+a_{2,3}\tau_2,~~~~~a_{3,3}
\end{array}\right),~~\tau=(\tau_1,\tau_2 )\in R^2.
\end{equation}
\begin{assumption}\label{a4} If $d=3$, then we assume that at least one of the following two conditions holds at each point $x^0\in \partial \mathcal O$: {\rm det}$B(x^0)\neq 1$ for unit vectors $\tau:|\tau|=1,$ or ${\rm Re}~a_{3,3}<0$. Note that the condition det$B\neq 1$ is equivalent to the non-degeneracy of the quadratic form with the matrix
$$
M(x^0):=\left ( \begin{array}{cc} a_{3,3}a_{1,1}-(a_{1,3})^2 & a_{3,3}a_{1,2}-a_{1,3}a_{2,3} \\
a_{3,3}a_{2,1}-a_{1,3}a_{2,3} & a_{3,3}a_{2,2}-(a_{2,3})^2  \end{array} \right )- I, \quad x^0\in \partial \mathcal O.
$$
If matrix $A$ is real, the latter condition simply means that {\rm det}$M>0$.
\end{assumption}

\textbf{Remarks.} 1. Obviously, assumption \ref{a4} holds if $\widetilde{A}$ is a diagonal matrix with positive elements $\lambda_1, \lambda_2,\lambda$ on the diagonal, $\lambda_1 \leq \lambda_2$ and $1/\lambda\notin [\lambda_1,\lambda_2]$.

2. Assumption \ref{a4} holds if $A(x^0)>I,~x_0 \in \partial \mathcal{O},$ or $0<A(x^0)<I,~x_0 \in \partial \mathcal{O},$ (the previous remark indicates that the latter inequalities are not necessary for the validity of the assumption \ref{a4}). To justify the remark, it is enough to note that {\rm det}$B(x^0)$ is a diagonal minor (the complement of the element $b_{1,1}$) of the matrix $\widehat{B}=D^*\widetilde{A}D$, where
$$
D=\left(
\begin{array}{l}
\tau_2~-\tau_1~~~0 \\
\tau_1~~~~\tau_2~~~~0 \\
0~~~~~~0~~~~1
\end{array}
\right).
$$
Hence, the eigenvalues of $B$ are located between the smallest and the largest of the eigenvalues of the matrix $\widehat{B}$. The latter eigenvalues  coincide with the eigenvalues of $A$.
\begin{proposition}\label{pro}
Let matrix $A$ be elliptic. Then Assumptions \ref{a3}, \ref{a4} are equivalent to the Shapiro-Lopatinskii condition for the ellipticity of the problem  (\ref{Anone}), (\ref{Antwo}).
\end{proposition}

\textbf{Proof.} We will assume that $d=3$. Let us apply the formal Fourier transform to (\ref{Anone1}), (\ref{Antwo1}) with respect to the tangential variables $y_1,y_2$.  Then the differentiation with respect to $y_1,y_2$ in (\ref{Anone1}), (\ref{Antwo1}) will be replaced by multiplication by  $i \tau_m, m=1,2$, and we obtain a boundary problem for the system of ordinary differential equations for functions $\widehat{u},\widehat{v}$ on the half line $t=y_3>0$. Consider only stable solutions of the corresponding equations, i.e., solutions vanishing at $t\to\infty$. The Shapiro-Lopatinskii condition (which defines the ellipticity of the problem (\ref{Anone}), (\ref{Antwo})) states that, for each $x^0\in \partial \mathcal O$, the homogeneous problem for $\widehat{u},\widehat{v}$  has only the trivial stable solution.

The following equation $\widehat{u}''(t)-\widehat{u}(t)|\tau|^2=0$ holds for $\widehat{u}$, and its stable solution is $\widehat{u}(t)=C_ue^{-t|\tau|}$, where $C_u=C_u(\tau)$ is a constant.
Equation for $\widehat{v}$ has the form
\begin{equation}\label{eqv}
\widehat{v}'' a_{3,3}+2i(a_{1,3}\tau_1+a_{2,3}\tau_2) \widehat{v}' -(\sum_{i,j=1}^2a_{i,j}\tau_i\tau_j)\widehat{v}=0,
\end{equation}
and, for each $\tau,~|\tau|\neq 0$, its characteristic equation
$$
\lambda^2 a_{3,3}+2i\lambda (a_{1,3}\tau_1+a_{2,3}\tau_2) \widehat{v}' -\sum_{i,j=1}^2a_{i,j}\tau_i\tau_j=0
$$
has exactly one root $\lambda=\lambda_0$ with a negative real part,
\begin{equation}\label{lambda0}
\lambda_0=\frac{ -(a_{1,3}\tau_1+a_{2,3}\tau_2)i-\sqrt{a_{3,3}\sum_{i,j=1}^2a_{i,j}\tau_i\tau_j-
(a_{1,3}\tau_1+a_{2,3}\tau_2)^2}}{a_{3,3}},
\end{equation}
due to the ellipticity of the matrices $A$ and $\widetilde{A}$. Note that the value under the root sign in the formula above equals det$B$. The value of the square root in that formula is chosen uniquely in such a way that
\begin{equation}\label{b}
{\rm Re}(\sqrt {{\rm det}B}/a_{3,3})>0.
\end{equation}
This choice is needed to guarantee that Re$\lambda_0<0$. The stable solution of (\ref{eqv}) is $\widehat{v}=C_ve^{\lambda_0t}$.

If $f=g=0$, the first boundary condition in (\ref{Antwo1}) implies that $C_u=C_v=c$, and from the second condition in (\ref{Antwo1}) it follows that
$$
c(-|\tau|-i(a_{1,3}\tau_1+a_{2,3}\tau_2)-a_{3,3}\lambda_0)=0.
$$
Hence, the ellipticity condition is
$$
-|\tau|-i(a_{1,3}\tau_1+a_{2,3}\tau_2)-a_{3,3}\lambda_0\neq 0 \quad \text{for} \quad |\tau|\neq 0,
$$
which is equivalent to
$$
\sqrt {\text{det }B}\neq |\tau| \quad \text{for} \quad |\tau|\neq 0.
$$
The latter condition can be written in the form $\sqrt {\text{det }B}\neq 1$ when $|\tau|=1$ since det$B$ is homogeneous in $\tau$. Thus, the Shapiro-Lopatinskii condition could be violated only if det$B=1$ for some $\tau,~|\tau|=1.$ However, it will not be violated even in this case if Re$a_{3,3}<0$ since the root $\sqrt {\text{det }B}$ is chosen by the condition (\ref{b}).

The proof is complete.

Consider the operator $L_k: H(\mathcal O) \rightarrow H(\mathcal O, \partial \mathcal O)$, where
$$
H(\mathcal O)=H^2(\mathcal O) \times H^2(\mathcal O), \quad H(\mathcal O, \partial \mathcal O)=H^0(\mathcal O) \times H^0(\mathcal O) \times H^{3/2}(\partial \mathcal O) \times H^{1/2}(\partial \mathcal O),
$$
which corresponds to the problem  (\ref{Anone}), (\ref{Antwo}) with non-zero right hand sides in all the equations, i.e.,
$$
L_k(u,v)=((\Delta+k^2) u, (\nabla A \nabla +k^2n) v, \gamma(u-v), \frac{\partial}{\partial \nu} u- \frac{\partial}{\partial \nu_A}v),
$$
where $\gamma$ is the trace operator and the last component of the vector on the right is also evaluated at the boundary. Recall that a linear operator in Hilbert spaces is called Fredholm if its range is closed and the dimensions of the kernel and co-kernel (orthogonal complement to the range) are finite and equal.

\begin{theorem}\label{mainT1}
Let the matrix $A$ be elliptic and Assumptions (\ref{a3}), (\ref{a4}) hold (i.e., the problem  (\ref{Anone}), (\ref{Antwo}) is elliptic). Then

a) there exists an integer $m\geq 0$ such the dimension of the kernel of the operator $L_k$ equals $m$ for all $k\in C$ except possibly a discrete set of points where the dimension of the kernel is finite, but greater than $m$; in particular, if $m=0$ at one point $k=k_0\in C$, then the interior transmission coefficients form a discrete set;

b) for all $k\in C$, the range of the operator $L_k$ is closed and the dimension $m_1$ of the co-kernel (the orthogonal complement to the range) is finite, the index  $\chi=m-m_1$ does not depend on $k$;

c) if the matrix $A$ is real valued, then $\chi=0$.
\end{theorem}

\textbf{Proof.} The first two statements of the theorem are the standard properties of the general elliptic boundary value problems which depend analytically on a parameter (see \cite{h},\cite{agr},\cite{grubb}). Let us prove the last one.  Since $\chi$ is $k$-independent, it is enough to show that $\chi=0$ for the operator $L_0$.

Consider the operator $S:H(\mathcal O) \rightarrow H(\mathcal O, \partial \mathcal O),$
$$
S(u,v)=(0,iv,0,0).
$$
The following inhomogeneous problem corresponds to the operator $L_0+\varepsilon S$
\begin{equation}\label{Anone2}
\begin{array}{l}
\Delta u =F\in H^2(\mathcal O), \quad x \in \mathcal O ,\\
\nabla A \nabla v + i\varepsilon v =G\in H^2(\mathcal O), \quad x \in \mathcal O,
\end{array}
\end{equation}
\begin{equation}\label{Antwo2}
\begin{array}{l}
u-v=f, \quad x \in \partial \mathcal O, \\
\frac{\partial u}{\partial \nu} - \frac{\partial v}{\partial \nu_A}=g, \quad x \in \partial \mathcal O.
\end{array}
\end{equation}
Let us show that the kernel of the operator
$L_0+\varepsilon S, ~\varepsilon>0,$ is trivial.

Let $(u,v) \in {\rm Ker}(L_0+\varepsilon S)$. The Green formulas imply
$$
0=\int_{\mathcal O}\left (\overline{u} \Delta u -   u\Delta \overline{u} \right )dx=\int_{\partial \mathcal O} \left ( \overline{u} \frac{\partial u}{\partial \nu}-u \overline{\frac{\partial u}{\partial \nu}}\right )dS,
$$
$$
0=\int_{\mathcal O} \left ( \overline{v}\nabla A \nabla v -v\overline{\nabla A \nabla v}-2i\varepsilon|v|^2\right )dx=-2i\varepsilon \int_{\mathcal O} |v|^2dx+\int_{\partial \mathcal O} \left ( \overline{v} \frac{\partial v}{\partial \nu_A}-v \overline{\frac{\partial v}{\partial \nu_A}}\right ).
$$
From here and the homogeneous boundary conditions (\ref{Antwo2}) for $u,v$ it follows that $v=0$. Since $u=v=0$ at the boundary and $u$ is harmonic in $\mathcal O$, function $u$ also vanishes in $\mathcal O$, i.e., the kernel of $L_0+\varepsilon S, ~\varepsilon>0,$ is trivial. The same arguments are valid if $\varepsilon<0.$

Let us find now the co-kernel of the operator $L_0+\varepsilon S, ~\varepsilon>0$. We look for the solution of the problem (\ref{Anone2}), (\ref{Antwo2}) in the form $(u,v)=(w,0)+(u_1,v_1),$ where $w\in H^2(\mathcal O)$ is an arbitrary function with $w=f,~\frac{\partial w}{\partial {\nu}}=g$ on $\partial \mathcal O$. This reduces the problem (\ref{Anone2}), (\ref{Antwo2}) to the problem for $(u_1,v_1)$ with different $F,G$ and homogeneous boundary conditions. The latter problem is solvable if the right-hand side in (\ref{Anone2}) is orthogonal to ${\rm Ker}(L_0-\varepsilon S)$, which is proved to be trivial. Hence, the problem (\ref{Anone2}), (\ref{Antwo2}) has a solution for arbitrary right-hand sides, i.e., the co-kernel of $L_0+\varepsilon S, ~\varepsilon>0,$ is trivial. Hence, $\chi=0$ for the operator $L_0+\varepsilon S, ~\varepsilon>0$. Since the index is homotopy invariant, $\chi=0$ for $L_0$.

The proof is complete.

The next couple of examples show that the interior transmission eigenvalues for an elliptic problem (\ref{Anone}), (\ref{Antwo}) may cover the whole complex plane, i.e., the kernel of an elliptic operator $L_k$ may be non-trivial for all $k\in C$.

\textbf{Two examples}. 1. Let $\mathcal O\in R^3$ be the unit cube, let $A$ be the diagonal matrix with elements $1,2,3$ on the diagonal, and let n(x)=1, i.e., the second equation in  (\ref{Anone}) has the form
$$
v_{x_1,x_1}+2v_{x_2,x_2}+3v_{x_3,x_3}+k^2u=0.
$$
Choose $u=v=a\cos kx_1+b\sin kx_1$. Then functions $(u,v)$ satisfy the homogeneous equations (\ref{Anone}), (\ref{Antwo}) for any $k\in C$, i.e., each $k\in C$ is an interior transmission eigenvalue, and the dimension $m$ of the kernel of the operator $L_k$ is at least two. On the other hand, the Shapiro-Lopatinskii condition holds on the smooth part of the boundary for the problem under consideration due to the Remark 1 after the Assumption \ref{a4}. The boundary $\partial\mathcal O$ is infinitely smooth in the next example.

2. Let $\mathcal O\in R^2$ be the disk $r<1$, where $(r,\phi)$ are polar coordinates in $R^2$, and let the second equation in (\ref{Anone}) be $\frac{\partial^2v}{\partial r^2} +\frac{1}{r}\frac{\partial v}{\partial r}+\frac{a(r)}{r^2}\frac{\partial^2v}{\partial \phi^2}+k^2v=0,$ where $a>0$ is a smooth positive function, $a(1)\neq 1$ and $a(r)=1$ for $r<1/2$. The above equation for $v$ is symmetric and can be written in the form indicated in (\ref{Anone}). The matrix $A$ will be smooth since the equation for $v$ coincides with the Laplacian when $r<1/2$. Since det$A(x_0)=a$ at the boundary, the assumptions of Theorem \ref{mainT1} hold. The dimension $m$ of the kernel of the operator $L_k$ in that example is at least one since $u=v$ in the kernel of the operator $L_k$ if they are smooth functions depending only on $r$ and satisfying the equation $y_{rr}+\frac{1}{r}y_r+k^2y=0.$ This example can be easily carried over to the spherical layer in any dimension.

In spite of these examples, the interior transmission coefficients form a discrete set in $C$ for elliptic problems with real matrix $A$ and generic potential $n$. To be more exact, the following statement holds.
\begin{theorem}\label{td}
Let the assumptions of Theorem \ref{mainT1} hold (i.e., problem (\ref{Anone}), (\ref{Antwo}) is elliptic) and let the matrix $A$ be real. Then there exists a finite set $Q$ of complex numbers such that the interior transmission eigenvalues form a discrete set in $C$ for any problem with $n(x)$ replaced by  $cn(x)$ if $c \notin Q$. The operator $L_k$ which corresponds to the problem with the potential $cn(x),~c \notin Q,$ is a  bijection when $k$ is not a transmission eigenvalue.
\end{theorem}
\textbf{Proof.} Denote by $L(k^2,k^2cn(x))$ the operator $L_k$ which corresponds to the problem with function $n(x)$ in the equation replaced by  $cn(x)$. The arguments $k^2,k^2cn(x)$ for the operator $L$ in the previous line indicate the coefficients for unknown functions in equations (\ref{Anone}). We also will need the operators $$
L^{(1)}=L(\kappa+i,\kappa^2n(x)),~L^{(2)}=L(\kappa^2+i,\kappa^2n(x)),
$$
which correspond to the problem (\ref{Anone}), (\ref{Antwo}), where the first equation is replaced by $\Delta u+(\kappa+i)u=0 $ and $\Delta u+(\kappa^2+i)u=0 $, respectively. After these changes of the first equation in (\ref{Anone}), the problem  (\ref{Anone}), (\ref{Antwo}) remains elliptic and depends analytically on the parameter $\kappa$. Hence, the first two statements of Theorem \ref{mainT1} are valid for the operators $L^{(1)}, L^{(2)}$. Moreover, it was shown in the process of the proof of the last statement of Theorem \ref{mainT1} that operators $L^{(1)},~L^{(2)}$ are bijections when $\kappa=0$. Thus there are discrete sets $\{\alpha_s,~s=1,2,...\}$ and $\{\beta_s,~s=1,2,...\}$ in $C$ such that the operators $L^{(j)},~j=1,2,$ are bijections when $\kappa\notin\{\alpha_s\}$ and $\kappa\notin\{\beta_s\}$, respectively.

Let us find now all the couples $(k,c)$ for which the operator $L(k^2,k^2cn(x))$ coincides with $L^{(1)}=L(\kappa+i,\kappa^2n(x))$ for some $\kappa\neq \alpha_s$. The operator $L(k^2,k^2cn(x))$ is a bijection for those $(k,c)$. This will happen when $k^2=\kappa+i,~~k^2c=\kappa^2,~~\kappa\neq \alpha_s,$ i.e., $k=\sqrt{\kappa+i},~c=\frac{\kappa^2}{\kappa+i}$. Hence, for each $c\neq \frac{\alpha_s^2}{\alpha_s+i}$, one can find a root $\kappa (c)$ of the quadratic equation $c=\frac{\kappa^2}{\kappa+i}$ and determine at least one value of $k=k(c)=\sqrt{\kappa(c)+i}$ for which $L(k^2,k^2cn(x))$ is a bijection. Then the operator $L(k^2,k^2cn(x))$ is a bijection for given $c$ and all $k\in C$ except, possibly, a discrete set of points, due to Theorem \ref{mainT1}. Thus the statement of Theorem \ref{td} is valid when $c\notin Q_1=\{\frac{\alpha_s^2}{\alpha_s+i},~s=1,2,...\}$.

Similarly, one can compare the operator  $L(k^2,k^2cn(x))$ and $L^{(2)}$. This will lead to the relations $k=\sqrt{\kappa^2+i},~c=\frac{\kappa^2}{\kappa^2+i},~\kappa\neq \beta_s$, which imply $c=\frac{\kappa^2}{\kappa^2+i}$. The latter equation can be solved for $\kappa$ if $c\neq 1$. Thus the statement of Theorem \ref{td} is valid when $c\notin Q_2=\{\frac{\beta_s^2}{\beta^2_s+i},~s=1,2,...\}\bigcup \{1\}$, and therefore it is valid when $c\notin Q=Q_1\bigcap Q_2$. It remains to note that the sets $\{\alpha_s\}$ and $\{\beta_s\}$ may have a limiting point only at infinity. Hence, $Q_1$ may have a limiting point only at infinity, $Q_2$ may have a limiting point only at $c=1$, and therefore their intersection $Q$ has at most a finite number of points.

The proof is complete.

The next statement provides conditions which guarantee that the set of the interior transmission eigenvalues is discrete for a chosen matrix $A$ and a potential $n$. These conditions consist of two parts:

First, we will assume that there exists a ray $l=\{e^{i\phi }\rho,~\rho\geq 0,\}$ in the complex plane for which the second equation in
(\ref{Anone}) is a parameter-elliptic equation with the parameter $k\in l$ (they are also called elliptic equations with a parameter, see \cite{agr}, \cite{grubb} and references there; not to be confused with elliptic equations depending on a parameter). This means the following. Consider the characteristic polynomial $P=P(x,\sigma,k)$ of the equation: $P=- \sigma\cdot A(x)\sigma +k^2n(x),~\sigma\in R^3,~k\in l.$ Recall that the ellipticity of the second equation in
(\ref{Anone}) (or the matrix $A$) implies that the quadratic (in $\sigma$) form $P(x,\sigma,0)=- \sigma\cdot A(x)\sigma$ is not degenerate.  The parameter-ellipticity means that the form  $P(x,\sigma,k)$, which is quadratic in $\sigma,~k\in l$, is not degenerate. Thus the first condition can be stated as follows.

\textit{Condition I.} There exists a constant $c>0$ such that $|P(x,\sigma,k)|\geq c(|\sigma|^2+|k|^2)$ for all $x\in \overline{\mathcal O},~\sigma\in R^3,~k\in l$.

Note that the latter inequality is homogeneous in $\sigma,~k\in l$ and can be checked only for $|\sigma|^2+|k|^2=1.$ Note also that the validity of the Condition I for some ray $l$ implies its validity for close rays. Hence, without loss of generality, we may assume that $l$ does not belong to the real axis, and therefore the first equation in (\ref{Anone}) is also parameter-elliptic when $k\in l$.

Condition I requires $n$ not to vanish, $n(x)\neq 0,~x\in \overline{\mathcal O}.$ The condition holds if the matrix $A$ is elliptic and real valued and $n(x)$ does not take values on some ray $l_1$ in the complex plane (for example, $n\neq 0$ is real, or $n\neq 0$ and Im $n\geq 0$, or $n\neq 0$ and Im $n\leq 0$). The condition also holds if an elliptic matrix $A$ is complex valued, the quadratic form $ \sigma\cdot A(x)\sigma,~x\in \overline{\mathcal O},~|\sigma|=1,$ does not take values on some ray $l_1$ and $n(x)\neq 0$ is real. Indeed, consider, for example, the second case with $n(x)>0$. Then condition I holds if $l$ is chosen in such a way that $k^2\in l_1$ when $k\in l$.

Condition II requires the whole boundary value problem (\ref{Anone}), (\ref{Antwo}) (not only equations (\ref{Anone})) to be parameter-elliptic when $k\in l$ \cite{agr}, \cite{grubb}. In order to state this condition in terms of $A$ and $n$ one needs to repeat the arguments used in the proof of Proposition \ref{pro} and write the Shapiro-Lopatinskii condition with the terms containing $k^2$ taken into account. The stable solution $\widehat{u}(t)$ will now have the form $\widehat{u}(t)=C_ue^{-t\sqrt{|\tau|^2-k^2}}$, where the root has to be chosen in such a way that Re$\sqrt{|\tau|^2-k^2}>0$ when $k\in l, ~|\tau|^2+|k|^2= 1.$ The stable solution $\widehat{v}(t)$ satisfies (\ref{eqv})  with an extra term $k^2n(x^0)\widehat{u}(t)$ on the left, i.e., $\widehat{v}(t)=C_ve^{-\lambda_kt}$, where
$$
\lambda_k=\frac{ -(a_{1,3}\tau_1+a_{2,3}\tau_2)i-\sqrt{a_{3,3}[\sum_{i,j=1}a_{i,j}\tau_i\tau_j-k^2n(x^0)]-
(a_{1,3}\tau_1+a_{2,3}\tau_2)^2}}{a_{3,3}}
$$
is the root of the characteristic polynomial of the equation for $\widehat{v}(t)$ such that Re$\lambda_k>0$ when $k\in l, ~|\tau|^2+|k|^2= 1.$ The existence of square roots with strictly positive real parts mentioned above is a consequence of Condition I.
Note that $\lambda_k$ with $k=0$ coincides with $\lambda_0$ defined in (\ref{lambda0}).

The first boundary condition in (\ref{Antwo1}) with $f=0$ implies that $C_u=C_v=c$. Then the second boundary condition with $g=0$ leads to
 $$
c(-\sqrt{|\tau|^2-k^2}-i(a_{1,3}\tau_1+a_{2,3}\tau_2)-a_{3,3}\lambda_k)=0.
$$
Boundary value problem (\ref{Anone}), (\ref{Antwo}) is parameter-elliptic if the latter relation implies that $c=0$, i.e., if
$$
\sqrt{|\tau|^2-k^2}\neq \sqrt{a_{3,3}[\sum_{i,j=1}^2a_{i,j}\tau_i\tau_j-k^2n(x^0) ]-(a_{1,3}\tau_1+a_{2,3}\tau_2)^2}, \quad k\in l, ~|\tau|^2+|k|^2= 1.
$$
This condition can be simplified by taking the square of both sides. Thus we assume that the following assumption holds.

\textit{Condition II.} There exists a constant $a>0$ such that
\begin{equation}\label{elp1}
|{\rm det}B-|\tau|^2-(a_{3,3}n(x^0)-1)k^2|\geq a(|\tau|^2+|k|^2)
\end{equation}
for all $x^0\in \partial \mathcal O,~\tau\in R^2,~k\in l$, where $l$ is the same ray as in Condition I (i.e., boundary value problem (\ref{Anone}), (\ref{Antwo}) is parameter-elliptic).
\begin{theorem}\label{last}
Let conditions I, II hold.
 Then the interior transmission eigenvalues form a discrete set, and the operator $L_k$ is bijective when $k$ is not an eigenvalue.
\end{theorem}

\begin{corollary}\label{con}
Let matrix $A$ be elliptic in $\overline{\mathcal O}$(see Definitions \ref{d1}, \ref{d2}), $n(x)\neq 0,~x\in \overline{\mathcal O},$ and $a_{d,d}n(x^0)\neq 1, ~x^0\in \partial\mathcal O$. Then conditions I,II hold and the statement of Theorem \ref{last} is valid in the following cases:

1) $A(x)$ and $n(x)$ are real, and ${\rm det}A(x^0)\neq 1,~x^0\in \partial\mathcal O,$ if $d=2,~~{\rm det}M(x^0)>0,~x^0\in \partial\mathcal O,$ if $d=3$ (see the Remarks following Assumption \ref{a4} for sufficient conditions);

or

2) the same conditions hold, but $n(x)$ is complex valued and there exists $\varepsilon>0$ such that $|{\rm Im}n(x)|>\varepsilon,~x\in \overline{\mathcal O}$;

or

3) both $A(x)$ and $n(x)$ are complex valued, Assumptions \ref{a3},  \ref{a4} hold, and there exists a ray $l$ in the complex plane which is free from the values of the following quadratic form $\frac{\sigma\cdot A(x)\sigma}{n(x)},~x\in \overline{\mathcal O},~|\sigma|=1,$  and the values of the function $\frac{{\rm det}A(x^0)-1}{a_{2,2}n(x^0)- 1},~x^0\in \partial\mathcal O,$ if $d=2$ or from the values of the following two quadratic forms
$$
\frac{\sigma\cdot A(x)\sigma}{n(x)},~x\in \overline{\mathcal O},~|\sigma|=1,~~{\it and}~~\frac{{\rm det}B(x^0,\tau)-1}{a_{3,3}n(x^0)- 1},~x^0\in \partial\mathcal O,~|\tau|=1,~~~{\it if} ~d=3.
$$

All the requirements of the case 3) are satisfied, for example, if the following three conditions are satisfied: a) $A(x)>0,~x\in \overline{\mathcal O},$ b) ${\rm det}A(x^0)>1,~x^0\in \partial\mathcal O,$ when $d=2$ or ${\rm det}B(x^0,\tau)>1~x^0\in \partial\mathcal O,~|\tau|=1,$ when $d=3$, and c) ${\rm Im}n(x)\geq 0,~x\in \overline{\mathcal O},$ or ${\rm Im}n(x)\leq 0,~x\in \overline{\mathcal O}$.
\end{corollary}



\textbf{Proof.} If a problem is parameter-elliptic, then the same problem with $k=0$ is elliptic, and therefore the assumptions of Theorem \ref{mainT1} hold if Conditions I, II are satisfied. The parameter-ellipticity of the problem  (\ref{Anone}), (\ref{Antwo}) implies that operator $L_k$ is bijective when $k\in l$ and $|k|$ is large enough \cite{agr}, \cite{grubb}. Thus the statement of Theorem \ref{last} follows from Theorem \ref{mainT1}. Let us prove the Corollary.

We will show first that Condition I holds in all the three cases mentioned in the Corollary, i.e., the quadratic (in $(\sigma,k)$) form $P(x,\sigma,k)=P(x,\sigma,0)+k^2n(x)$ is non-degenerate for all $x \in \overline{\mathcal O}$. In the first case, $P(x,\sigma,0)$ and $n(x)$ are real and do not vanish if $|\sigma|\neq 0$ (due to the ellipticity of the matrix $A$ and the condition imposed on $n$). Then the sum $P(x,\sigma,k)$ is a non-degenerate form when $k\in l$ for an arbitrary ray $l$ which does not belong to the real axis. In the second case ($A$ is real, $|{\rm Im }n(x)|> \varepsilon$), one can choose any ray $l$ with small enough polar angle $\phi$. Indeed, let us note that $ |n(x)|\leq c_0<\infty $, i.e., the polar angles $\theta\in[0,2\pi)$ of the complex numbers $n(x),~x\in \overline{\mathcal O},$ are separated from $0,\pi$ and $2\pi$. Then $k^2n(x)$ is not real when $0\neq k\in l$ and $|\phi|$ is small enough. Since $P(x,\sigma,0)$ is real and does not vanish when $|\sigma|\neq 0$, the form $P(x,\sigma,k)=P(x,\sigma,0)+k^2n(x)$ does not vanish when $k\in l,~|\sigma|^2+|k|^2=1$. Thus Condition I holds in the second case of the Corollary. Similarly, in the third case, the form $-\frac{\sigma\cdot A(x)\sigma}{n(x)}+k^2$ is non-degenerate when $k\in l$ and therefore Condition I holds.

The validity of Condition II in all the three cases can be justified absolutely similarly since it also requires the non-degeneracy of the sum of two forms: a quadratic form in $\tau$ and the form $(a_{3,3}n(x^0)-1)k^2$.

\textbf{Acknowledgment.} The authors are very grateful to F. Cakoni, D. Colton and A. Kirsch for useful discussions.

\end{document}